\documentclass[12pt,a4paper]{article}
\newcommand{\be}{\begin{equation}}
\newcommand{\ee}{\end{equation}}
\newcommand{\bea}{\begin{eqnarray}}
\newcommand{\eea}{\end{eqnarray}}
\newcommand{\ba}{\begin{array}}
\newcommand{\ea}{\end{array}}
\newcommand{\bean}{\begin{eqnarray*}}
\newcommand{\eean}{\end{eqnarray*}}
\newcommand{\no}{\nonumber}
\newcommand{\Th}{\Theta}
\newcommand{\bth}{\bar{\theta}}
\newcommand{\tth}{\tilde{\theta}}
\newcommand{\tTh}{\tilde{\Theta}}
\newcommand{\hTh}{\hat{\Theta}}

\newcommand{\la}{\lambda}
\newcommand{\bla}{\bar{\lambda}}
\newcommand{\bmu}{\bar{\mu}}
\newcommand{\bn}{\bar{n}}
\newcommand{\bm}{\bar{m}}
\newcommand{\bu}{\bar{u}}

\newcommand{\mB}{{\mathcal B\/}}
\newcommand{\mF}{{\mathcal F\/}}
\newcommand{\bmB}{\bar{{\mathcal B\/}}}

\newcommand{\bB}{\bar{{\mathcal B\/}}}

\newcommand{\bQ}{\bar{Q}}
\newcommand{\bD}{\bar{D}}
\newcommand{\bS}{\bar{S}}

\newcommand{\bt}{\bar{t}}
\newcommand{\bp}{\bar{p}}
\newcommand{\brf}{\bar{f}}

\newcommand{\pa}{\partial}

\newcommand{\res}{{\rm res\/}}

\newcommand{\fixed}{{\rm fixed\/}}
\newcommand{\wbar}{\overline}
\begin{document}
\baselineskip=0.7cm
\title
     {\bf On Kernel Formulas and Dispersionless Hirota Equations\/}
\author
{\sc Yu-Tung Chen$^1$\footnote{E-mail: ytchen@math.sinica.edu.tw}
 and Ming-Hsien Tu$^2$\footnote{E-mail: phymhtu@ccu.edu.tw(corresponding author)} \\ \\
  $^1$%%
  {\it Institute of Mathematics, Academia Sinica, Taipei, Taiwan\/},\\
   $^2$%%
  {\it Department of Physics, National Chung Cheng University\/},\\
   {\it Minghsiung, Chiayi 621, Taiwan\/}\/}
\date{\today}
  \maketitle
  \begin{abstract}
 We rederive dispersionless Hirota equations of the dispersionless Toda hierarchy
 from the method of kernel formula  provided by Carroll and Kodama.
  We then apply the  method to derive  dispersionless Hirota equations
  of  the extended  dispersionless  BKP(EdBKP) hierarchy proposed by Takasaki.
  Moreover, we verify associativity equations (WDVV equations) in the EdBKP hierarchy
  from  dispersionless Hirota equations and give a realization of associative algebra
  with structure constants expressed in terms of residue formula.
\end{abstract}
\newpage
\section{Introduction}

Dispersionless integrable systems (DIS) are integrable hierarchies in
dispersionless limit (or quasi-classical limit).
Among them the dispersionless Kadomtsev-Petviashvili (dKP) and  dispersionless
Toda (dToda) hierarchies are special since they have been recognized as universal DIS
in several fields of theoretical physics and mathematics
( see e.g. \cite{AK96,BK00,D,Dub96,EGLS,Kod88a,KG89,Kri92,Li99,TT92,TT95} and references therein).
The solutions of the dKP and dToda hierarchies
can be characterized by a single function called $\tau$ functions, the logarithm
of them; namely, $\mF=\log\tau$ describe free energy of some two-dimensional
 topological field theories in genus zero and satisfy the
 Witten-Dijkgraaf-Verlinde-Verlinde (WDVV) equations\cite{W,DVV91,Dub96}. In particular,
the finite-dimensional reductions of dKP and dToda systems are realized as
special solutions of topological Landau-Ginzburg models of $A$-type and
topological $CP^1$ models, respectively\cite{DVV91,Kri92,EY94,AK96}. In terms of free energy $\mF$,
the hierarchy flows of dKP and dToda can be written as a set of second derivatives of
$\mF$ called dispersionless Hirota (dHirota) equations.
There are a lot of works devoted to derive dHirota equations.
In \cite{TT95}, Takasaki and Takebe derived dHirota equations of
the dKP hierarchy from dispersionless limit of differential Fay identity.
Carroll and Kodama \cite{CK95} studied the same dHirota equation from the method
of kernel formula. On the other hand, Wiegmann, Zabrodin {\it et al\/} \cite{KKMWZ,WZ}
investigated dHirota equations of the dToda hierarchy in the context of conformal mapping.
More recently, Teo \cite{Teo03a} derived  dHirota equations of the dKP and dToda hierarchies
from complex analysis using the notions of Grunsky coefficients and Faber polynomials.

The main purpose of the work is to demonstrate the applicability of kernel formula to
other dispersionless integrable hierarchies.
In particular we would like to derive dHirota equations of a universal integrable hierarchy
underlying topological Landau-Ginzburg models of $D$-type proposed by Takasaki \cite{T93b}.
Since this integrable hierarchy is an extension of the dispersionless BKP (dBKP)
hierarchy\cite{DR92,T93a,KA02,BK}
and has two sets of time variables we refer it to the extended dispersionless BKP (EdBKP)
hierarchy \cite{ChT}.  The EdBKP hierarchy resembles the dToda hierarchy in many formulations
such as dressing operators, Orlov functions,
Riemann-Hilbert problem, additional symmetries, $w$-algebras, hodograph solutions
etc.\cite{T93b,ChT}  Motivated by the work for dKP\cite{CK95} we first give an elementary
derivation of dHirota equations to the dToda hierarchy by using the method of kernel formula.
We then show that the method of kernel formula can be applied to the EdBKP hierarchy without
difficulty.
As a by-product, associativity equations (WDVV equations) in the EdBKP hierarchy can be
 verified directly from dHirota equations.
Therefore our results indeed provide another point of view to understand
the integrability of the dBKP hierarchy in connection to topological field theories.

Let us briefly recall the  method of kernel formula\cite{CK95} for the derivation of
dHirota equations of  the dKP hierarchy. The dKP hierarchy is defined by the Lax equations
\[
\pa_{t_n}\la=\{\mB_n,\la\},\quad \mB_n=(\la^n)_{\geq 0},\quad n=1,2,\cdots
\]
where the Lax operator $\la$ is a Laurent series of the form
\[
\la=p+\sum_{n=1}u_{n+1}p^{-n},
\]
with $u_j$ are functions of the time variables $t=(x=t_1,t_2,\cdots)$, $(A(p))_{\geq 0}$ denotes
the projection on the polynomial part of $A(p)$, and $\{A, B\}$ stands for the Poisson bracket
\[
\{A,B\}=\frac{\pa A}{\pa p}\frac{\pa B}{\pa x}-\frac{\pa A}{\pa x}\frac{\pa B}{\pa p}.
\]
It can be shown\cite{TT92,TT95} that there exists a function $\mF(t)$ called free energy such that
\[
\mB_n(\la)=\la^n-\sum_{m=1}\frac{\mF_{n,m}}{m}\la^{-m}
\]
where $\mF_{n,m}=\pa_{t_n}\pa_{t_m}\mF$ and thus $\mF_{n,m}=\mF_{m,n}$.
Particularly, for $\mB_1=p$ we have
\be
p(\la)=\la-\sum_{m=1}\frac{\mF_{1,m}}{m}\la^{-m}
\label{B1}
\ee
which can be viewed as the inverse map of $\la(p)$.
Multiplying $\la^{n-1}\pa_p\la$ with $n\geq 1$ to (\ref{B1}) and taking the projection
$()_{\geq 0}$ we obtain
\[
\pa_pQ_{n+1}(\la)-p(\la)\pa_pQ_n(\la)-\sum_{m=1}^{n-1}\frac{\mF_{1,m}}{m}\pa_pQ_{n-m}(\la),\quad n\geq 1.
\]
where $Q_n=\mB_n/n$. Multiplying $\mu^{-n}$ and summing over $n\geq 0$ we have
\bean
1&=&(\mu-p(\la)-\sum_{m=1}\frac{\mF_{1,m}}{m}\mu^{-m})\sum_{j=1}\pa_pQ_j(\la)\mu^{-j},\\
&=&(p(\mu)-p(\la))\sum_{j=1}\pa_pQ_j(\la)\mu^{-j},
\eean
or
\[
\frac{1}{p(\mu)-p(\la)}=\sum_{j=1}^\infty\pa_pQ_j(\la)\mu^{-j}
\]
which is the kernel formula for the generating function of $\pa_pQ_j(\la)$.
Integrating above with respect to $p(\la)$ and fixing the normalization at $\mu=\infty$, then
\[
\frac{\mu}{p(\mu)-p(\la)}=e^{\sum^\infty_{j=1}Q_j(\la)\mu^{-j}}
\]
which together with the expressions of second derivatives of $\mF$
for $\mB_j(\la)$ and $p(\la)$ yields
 dHirota equations of the dKP hierarchy \cite{TT95,CK95}:
\be
D(\mu)\pa_{t_1}\mF-D(\la)\pa_{t_1}\mF=(\mu-\la)(1-e^{D(\mu)D(\la)\mF})
\label{dH-dKP}
\ee
where $D(z)=\sum_{m=1}^\infty z^{-m}\pa_{t_m}/m$. A convenient way to obtain
 a set of relations defined by $\mF_{m,n}$ is to consider (\ref{dH-dKP}) under
 the limit $\mu\to \la$. Then one gets
 \[
\pa_{\la}D(\la)\pa_{t_1}\mF=1-e^{D(\la)D(\la)\mF}.
 \]
By comparing the coefficients of $\la^{-n}$ on both sides one obtains \cite{CK95}:
 \[
\mF_{1j}=P_{j+1}(Z_1=0,Z_2,\cdots,Z_{j+1}),\quad j\geq 1
 \]
 where $P_j$ are elementary Schur polynomials and $Z_j=\sum_{n+m=j}\frac{\mF_{nm}}{nm}$.
 A few equations from such an expansion have been given in \cite{CK95}. Here we just
 mention the simplest nontrivial equation
\[
\frac{1}{2}\mF^2_{11}-\frac{1}{3}\mF_{13}+\frac{1}{4}\mF_{22}=0
\]
which, after noticing $u=u_2=\mF_{11}$, is just the $(2+1)$-dimensional dKP
equation\cite{Kod88a}
\[
u_t=3uu_x+\frac{3}{4}\pa_x^{-1}u_{yy}.
\]
Therefore, kernel formula provides an elementary and convenient way to derive
dHirota equations of dKP from the Lax formulation without referring to its
dispersive counter parts, such as (differential) Fay identity.
The rest of the  paper is organized as follows. In sec. 2 we rederive dHirota
equations of the dToda hierarchy using the method of kernel formula.
In sec. 3 we turn to the EdBKP hierarchy to investigate its dispersionless Hirota equations.
In sec. 4 we verify associativity equations (or WDVV equations) in the EdBKP hierarchy
from dispersionless Hirota equations. Realization of associative algebra and the residue
formula for structure constants of the associative algebra are also given.
Sec. 5 is devoted to the concluding remarks.
%%%%%%%%%%%%%%%%%%%%%%%%%%%%%%%%%%%%%%%%%%%%%%%%%%%%%%%%%%%%%%%%%%%%%%%%%
\section{Dispersionless Toda hierarchy}
%%%%%%%%%%%%%%%%%%%%%%%%%%%%%%%%%%%%%%%%%%%%%%%%%%%%%%%%%%%%%%%%%%%%%%%%%%
Next let us apply the method of kernel formula to the dToda hierarchy as a warm-up.
Following \cite{TT91,TT95}, the dToda hierarchy has two series of independent variables
$t=(t_1,t_2,\cdots)$ and $\bt=(\bt_1,\bt_2,\cdots)$
along with a space variable $t_0$. Its Lax representation is
\bean
\pa_{t_n}\la=\{\mB_n,\la\},\quad \pa_{\bt_n}\la=\{\bB_n,\la\}, \\
\pa_{t_n}\bla=\{\mB_n,\bla\},\quad \pa_{\bt_n}\bla=\{\bB_n,\bla\},
\eean
where $\la$ and $\bla$ are Laurent series
\[
\la=p+\sum_{n=0}u_{n+1}p^{-n},\quad \bla^{-1}=\bar{u}_0p^{-1}+\sum_{n=0}\bar{u}_{n+1}p^n
\]
of a variable $p$ and $\mB_n$ and $\bB_n$ are given by
\[
\mB_n=(\la^n)_{\geq 0},\quad \bB_n=(\bla^{-n})_{\leq -1},
\]
where we denote $(A)_{\leq -1}=A-(A)_{\geq 0}$.
The Poisson bracket $\{ , \}$ is defined by
\[
\{A(p,t),B(p,t)\}=p\frac{\pa A(p,t)}{\pa p}\frac{\pa B(p,t)}{\pa t_0}-
p\frac{\pa A(p,t)}{\pa t_0}\frac{\pa B(p,t)}{\pa p}.
\]
One can view $\la$ as a  map defined in domain $D_\infty$ containing $p=\infty$,
while $\bla$ in $D_0$ containing $p=0$.
Let $\mF(t_0,t,\bt)$ be the free energy associated with the dToda hierarchy,
one can define functions $S$ and $\bS$ \cite{TT91,TT95}:
\bean
S(\la)&=&\sum_{n=1}t_n\la^n+t_0\log\la-\sum_{n=1}^\infty\frac{\pa_{t_n}\mF}{n}\la^{-n},\\
\bS(\bla)&=&\sum_{n=1}\bt_n\bla^{-n}+t_0\log\bla+\phi-\sum_{n=1}^\infty\frac{\pa_{\bt_n}\mF}{n}\bla^{n},
\eean
such that
\bea
\label{B}
\mB_n(\la)&=&\pa_{t_n}S(\la)=\la^n-\sum_{m=1}\frac{\mF_{n,m}}{m}\la^{-m},\\
\label{Bb}
\mB_n(\bla)&=&\pa_{t_n}\bS(\bla)=\mF_{n0}-\sum_{m=1}\frac{\mF_{n,\bm}}{m}\bla^m,\\
\label{bB}
\bB_n(\la)&=&\pa_{\bt_n}S(\la)=-\sum_{m=1}\frac{\mF_{\bn,m}}{m}\la^{-m},\\
\label{bBb}
 \bB_n(\bla)&=&\pa_{\bt_n}\bS(\bla)=\bla^{-n}+\mF_{0,\bn}-\sum_{m=1}\frac{\mF_{\bn,\bm}}{m}\bla^m,\\
 \label{lp}
\log p(\la)&=&\pa_{t_0}S(\la)=\log\la-\sum_{m=1}\frac{\mF_{0,m}}{m}\la^{-m},\\
\label{lpb}
\log p(\bla)&=&\pa_{t_0}\bS(\bla)=\log\bla+\mF_{0,0}-\sum_{m=1}\frac{\mF_{0,\bm}}{m}\bla^m,
\eea
where $\mF_{n,\bm}=\pa_{t_n}\pa_{\bt_m}\mF$ and $\phi$ satisfies
\[
\pa_{t_n}\phi=(\la^n)_0,\quad \pa_{\bt_n}\phi=-(\bla^{-n})_0,\quad \pa_{t_0}\phi=\log \bu_0.
\]
Noticing that (\ref{B}), (\ref{bB}), and (\ref{lp}) are defined in the domain $D_\infty$,
 while (\ref{Bb}), (\ref{bBb}), and (\ref{lpb}) in $D_0$.

In contrast to the  dKP hierarchy where only one Lax operator is defined in $D_\infty$,
 the dToda hierarchy contains two Lax operators $\la$ and $\bla$ defined in domains $D_\infty$
 and $D_0$, respectively. To derive dispersionless Hirota equations for the dToda hierarchy
we shall generalize the method of kernel formula to the domains $D_\infty$, $D_0$,
and $D_\infty \cap D_0$.

In $D_\infty$. From equation  (\ref{B}) for $n=1$ we have
\[
p(\la)+u_1=\la-\sum_{m=1}^\infty\frac{\mF_{1,m}}{m}\la^{-m}.
\]
Multiplying both-hand sides by $\la^{n-1}\pa_p\la$ with $n\geq 1$ and taking the projection $()_{\geq 0}$
one gets
\[
\pa_p Q_{n+1}(\la)-(p(\la)+u_1)\pa_p Q_n(\la)-\sum_{m=1}^{n-1}\frac{\mF_{m,1}}{m}\pa_p Q_{n-m}(\la)=0,\quad n\geq 1
\]
where $Q_n=\mB_n/n$. Multiplying $\mu^{-n}$ and summing over $n\geq 0$ we have
\bean
1&=&\sum_{n=0}^\infty(\pa_p Q_{n+1}(\la)-(p(\la)+u_1)\pa_p Q_n(\la)-
\sum_{m=1}^{n-1}\frac{\mF_{m,1}}{m}\pa_p Q_{n-m}(\la))\mu^{-n},\\
&=&(\mu-u_1-\sum_{j=1}^\infty\frac{\mF_{1,j}}{j}\mu^{-j}-p(\la))\sum_{m=1}^\infty\pa_p Q_m(\la)\mu^{-m}\\
&=&(p(\mu)-p(\la))\sum_{m=1}^\infty\pa_p Q_m(\la)\mu^{-m}
\eean
or
\be
\frac{1}{p(\mu)-p(\la)}=\sum_{m=1}^\infty\pa_p Q_m(\la)\mu^{-m}.
\label{kf1}
\ee
Integrating the kernel formula (\ref{kf1}) with respect to $p(\la)$ and fixing
the normalization at $\mu=\infty$ we obtain
\[
p(\mu)-p(\la)=\mu e^{-\sum_{m=1}^\infty Q_m(\la)\mu^{-m}}
\]
which, after substituting the expressions of $Q_n(\la)$ and $p(\la)$, becomes
\be
\mu e^{-D(\mu)\pa_{t_0}\mF}-\la e^{-D(\la)\pa_{t_0}\mF}=(\mu-\la)e^{D(\la)D(\mu)\mF}.
\label{dH1}
\ee
where $D(z)=\sum_{m=1}^\infty z^{-m}\pa_{t_m}/m$.

In $D_0$. Starting from equation (\ref{bBb}) for $n=1$, we have
\[
\bu_0p^{-1}(\bla)=\bla^{-1}+\mF_{0,\bar{1}}-\sum_{m=1}^\infty\frac{\mF_{\bar{1},\bm}}{m}\bla^m
\]
Multiplying both-hand sides by $\bla^{-n-1}\pa_p\bla$ with $n\geq 1$ and
taking the projection $()_{\leq -2}$ one gets
\[
\bu_0p^{-1}(\bla)\pa_p\bQ_n-\pa_p\bQ_{n+1}-\mF_{0,\bar{1}}\pa_p\bQ_n
+\sum_{m=1}^{n-1}\frac{\mF_{\bar{1},\bm}}{m}\pa_p\bQ_{n-m}=0,\quad n\geq 1.
\]
where $\bQ_n=\bmB_n/n$. Multiplying above by $\bmu^{n}$ and sum over $n\geq 0$ we have
\be
\frac{p(\bmu)}{p(\bla)(p(\bmu)-p(\bla))}=\sum_{m=1}^\infty\pa_p\bQ_m(\bla)\bmu^m
\label{kf2}
\ee
which is the kernel formula for the generating function of $\pa_p\bQ_m(\bla)$.
Integrating above with respect to $p(\bla)$ and fixing the normalization at $\bmu=0$
we obtain
\[
\frac{p(\bla)-p(\bmu)}{p(\bla)}=\frac{\bla-\bmu}{\bla}
e^{-\bD(\bmu^{-1})\pa_{t_0}\mF+\bD(\bla^{-1})\bD(\bmu^{-1})\mF}
\]
which, after substituting the expression of $\log p(\bla)$, becomes
\be
\bla e^{\bD(\bmu^{-1})\pa_{t_0}\mF}-\bmu e^{\bD(\bla^{-1})\pa_{t_0}\mF}=
(\bla-\bmu)e^{\bD(\bla^{-1})\bD(\bmu^{-1})\mF}
\label{dH2}
\ee
where $\bD(z)=\sum_{m=1}^\infty z^{-m}\pa_{\bt_m}/m$.

In $D_\infty \cap D_0$. By replacing $Q_m(\la)$ and $p(\la)$  in (\ref{kf1}) by $Q_m(\bla)$ and
$p(\bla)$, respectively we get
\[
\frac{1}{p(\mu)-p(\bla)}=\sum_{m=1}^\infty\pa_p Q_m(\bla)\mu^{-m}
\]
where $Q_m(\bla)=B_m(\bla)/m$. Integrating above with respect to $p(\bla)$ and
fixing the normalization at $\mu=\infty$ we obtain
\[
\frac{p(\mu)-p(\bla)}{\mu}=e^{-D(\mu)\pa_{t_0}\mF+\bD(\bla^{-1})D(\mu)\mF}
\]
Substituting the expression of $\log p(\bla)$ and $\log p(\la)$ into above we get
\be
1-\frac{\bla}{\mu}e^{\mF_{0,0}+D(\mu)\pa_{t_0}\mF-\bD(\bla^{-1})\pa_{t_0}\mF}=e^{D(\mu)\bD(\bla^{-1})\mF}.
\label{dH3}
\ee
Similarly, we can also consider (\ref{kf2}) in  $D_\infty \cap D_0$ by replacing
  $\bQ_m(\bla)$ and $p(\bla)$ by $\bQ_m(\la)$ and $p(\la)$, respectively. Then
\[
\frac{p(\bmu)}{p(\la)(p(\bmu)-p(\la))}=\sum_{m=1}^\infty\pa_p \bQ_m(\la)\bmu^m
\]
 Integrating above with respect to $p(\la)$ and fixing the normalization at $\bmu=0$ we obtain
\[
\frac{p(\la)-p(\bmu)}{p(\la)}=e^{-\sum_{m=1}^\infty\bQ_m(\la)\bmu^m}
\]
which together with (\ref{lp}) and (\ref{lpb}) implies
\[
1-\frac{\bmu}{\la}e^{\mF_{0,0}+D(\la)\pa_{t_0}\mF-\bD(\bmu^{-1})\pa_{t_0}\mF}=
e^{D(\la)\bD(\bmu^{-1})\mF}.
\]
This equation, after making a replacement $\la\to \mu$ and $\bmu\to\bla$,
is the same as (\ref{dH3}). In summary, the free energy $\mF$ of the dToda hierarchy
is characterized by the dHirota equations ((\ref{dH1}), (\ref{dH2}),
and (\ref{dH3}))\cite{WZ,KKMWZ,Z,Teo03a}:
\bea
&&\mu e^{-D(\mu)\pa_{t_0}\mF}-\la e^{-D(\la)\pa_{t_0}\mF}=(\mu-\la)e^{D(\la)D(\mu)\mF},\no\\
&&e^{\mF_{0,0}+D(\mu)\pa_{t_0}\mF-\bD(\la)\pa_{t_0}\mF}=\mu\la(1-e^{D(\mu)\bD(\la)\mF}),\no\\
&&\mu e^{\bD(\mu)\pa_{t_0}\mF}-\la e^{\bD(\la)\pa_{t_0}\mF}=
(\mu-\la)e^{\bD(\la)\bD(\mu)\mF}
\label{dH-toda}
\eea
where  we have made a change of variables $\bla\to \la^{-1}$ and $\bmu\to \mu^{-1}$
in (\ref{dH2}) and (\ref{dH3}) for convenience to extract a set of relations
satisfied by the second derivatives of $\mF$.
For the first equation in (\ref{dH-toda}), we take the limit $\la\rightarrow\mu$ to obtain
\[
 \pa_{\mu}(\mu e^{-D(\mu)\pa_{t_0}\mF}) = e^{D^2(\mu)\mF},
\]
which, in terms of the second derivatives of $\mF$, becomes
\[
 1+\sum_{k=1}^{\infty}\mF_{0k}\mu^{-k}
 =\exp\left(\sum_{m=1}^{\infty}\frac{\mF_{0m}}{m}\mu^{-m}
  +\sum_{m,n=1}^{\infty}\frac{\mF_{mn}}{mn}\mu^{-m-n}\right).
\]
Using the elementary Schur polynomial and eliminating the constant terms on both sides,
we obtain a set of dHirota equations of dToda hierarchy in terms of
$t_0$- and $t_k$-derivatives of $\mF$:
\be
 \mF_{0k}-P_k(\Th_1,\Th_2,\ldots,\Th_k)=0,
  \label{dTodaH-1}
\ee
where
\[
 \Th_1 = \mF_{01}, \quad  \Th_i = \frac{\mF_{0i}}{i} +
          \sum_{\scriptstyle m+n=i \atop\scriptstyle m,n\geq 1}\frac{\mF_{mn}}{mn},
          \quad i\geq 2.
\]
In a similar way, from the third equation in (\ref{dH-toda}),
we have another set of dHirota equations in terms of
$t_0$- and $\bt_k$-derivatives of $\mF$:
\be
  \mF_{0\bar{k}}+P_k(\bar{\Th}_1,\bar{\Th}_2,\ldots,\bar{\Th}_k)=0,
 \label{dTodaH-2}
\ee
where
\[
 \bar{\Th}_1 = -\mF_{0\bar{1}}, \quad  \bar{\Th}_i = -\frac{\mF_{0\bar{i}}}{i} +
          \sum_{\scriptstyle m+n=i \atop\scriptstyle m,n\geq 1}\frac{\mF_{\bar{m}\bar{n}}}{mn},
          \quad i\geq 2.
\]
For the second equation in (\ref{dH-toda}), by taking the limit $\la\rightarrow\mu$, we have
\[
1-\frac{1}{\mu^2}e^{\mF_{00}}\sum_{k=0}^{\infty}P_k(\tTh_1,\tTh_2,\ldots,\tTh_k)\mu^{-k}
= \sum_{k=0}^{\infty}P_k(\hTh_1=0,\hTh_2,\ldots,\hTh_k)\mu^{-k},
\]
where $P_k(\tTh)$ and $P_k(\hTh)$ are the elementary Schur polynomials and
\[
 \tTh_i = \frac{1}{i}(\mF_{0i}-\mF_{0\bar{i}}), \quad i\geq 1, \quad
 \hTh_i = \sum_{\scriptstyle m+n=i \atop\scriptstyle m,n\geq 1}\frac{\mF_{m\bar{n}}}{mn},
            \quad i\geq 2.
\]
By eliminating the constant terms on both sides and noticing $P_1(\hTh)=0$,
we get a set of dHirota equations  involving
$t_0$-, $t_k$- and $\bt_k$-derivatives of $\mF$:
\be
 e^{\mF_{00}}P_k(\tTh)+P_{k+2}(\hTh) = 0.
  \label{dTodaH-3}
\ee
 Since we don't find equations from such expansions in the literature,
 we list a few of them in appendix A.
  In particular, the first nontrivial equation yields
\[
e^{\mF_{00}}+\mF_{1\bar{1}}=0,
\]
which, after differentiating with respect to $t_0$ and noticing $\phi=\pa_0\mF$,
is just the dToda equation \cite{TT91,TT93,TT95}
\[
\pa_{t_1}\pa_{\bar{t}_1}\phi+\pa_{t_0}e^{\pa_{t_0}\phi}=0.
\]

%%%%%%%%%%%%%%%%%%%%%%%%%%%%%%%%%%%%%%%%%%%%%%%%%%%%%%%%
\section{Extended dispersionless BKP hierarchy}
%%%%%%%%%%%%%%%%%%%%%%%%%%%%%%%%%%%%%%%%%%%%%%%%%%%%%%%%
\subsection{Lax formalism}
%%%%%%%%%%%%%%%%%%%%%%%%%%%%%%%%%%%%%%%%%%%%%%%%%%%%%%%%%%%%%%%%
Having derived dHirota equations of the dToda hierarchy, now we
like to apply the method of kernel formula to the EdBKP hierarchy  which is the
integrable hierarchy underlying topological Landau-Ginzburg models of $D$-type
proposed by Takasaki\cite{T93b}(see also \cite{ChT}). The EdBKP hierarchy is described by
\bea
&&\pa_{t_{2n+1}}\la=\{\mB_{2n+1},\la\},
\quad \pa_{\bt_{2n+1}}\la=\{\bmB_{2n+1},\la\},  \no\\
&&\pa_{t_{2n+1}}\bla=\{\mB_{2n+1},\bla\},
\quad \pa_{\bt_{2n+1}}\bla=\{\bmB_{2n+1},\bla\},\quad n=0,1,2,\ldots
\label{Leq}
\eea
with
\[
\la=p+\sum_{n=1}^\infty u_{2n}p^{-2n+1},\quad
\bla=\sum_{n=0}^\infty\bar{u}_{2n}p^{2n+1},\quad \bar{u}_0\neq 0
\]
and
\[
\mB_{2n+1}=(\la^{2n+1})_{\geq 0},\quad \bmB_{2n+1}=(\bla^{-2n-1})_{\leq -1},
\]
where the coefficient functions $u_{2n}$ and $\bar{u}_{2n}$ depend on
the time variables $t=(x=t_1,t_3,\ldots)$ and $\bar{t}=(\bar{t}_1,\bar{t}_3,\ldots)$ and
the Poisson bracket $\{,\}$ here is defined by
\[
 \{f,g\}=\frac{\pa f}{\pa p}\frac{\pa g}{\pa x} - \frac{\pa f}{\pa x}\frac{\pa g}{\pa p}.
\]
Noting that $\la(-p)=-\la(p)$ and $\bla(-p)=-\bla(p)$ and thus
$(\mB_{2n+1})_{[0]}=(\bmB_{2n+1})_{[0]}=0$.
The Lax equations (\ref{Leq}) are equivalent to the zero curvature equations
\bean
&&\pa_{t_{2n+1}}\mB_{2m+1}-\pa_{t_{2m+1}}\mB_{2n+1}+
\{\mB_{2m+1},\mB_{2n+1}\}=0,\no\\
&&\pa_{\bt_{2n+1}}\bmB_{2m+1}-\pa_{\bt_{2m+1}}\bmB_{2n+1}+
\{\bmB_{2m+1},\bmB_{2n+1}\}=0,\no\\
&&\pa_{\bt_{2n+1}}\mB_{2m+1}-\pa_{t_{2m+1}}\bmB_{2n+1}+
\{\mB_{2m+1},\bmB_{2n+1}\}=0,
\label{zero}
\eean
which guarantees that the Lax equations (\ref{Leq}) commute between themselves.
Since the first member involving
$t_{2n+1}$-flows only is just the dBKP hierarchy and that is the reason why we call
this integrable hierarchy the extended dBKP hierarchy.
It can be shown\cite{T93b} that there exists a single function $\mF(t,\bt)$
(free energy) from which one can define the functions
\bean
S(\la) &=& \sum_{n=0}^{\infty}t_{2n+1}\la^{2n+1}
          -\sum_{n=0}\frac{1}{2n+1}\pa_{t_{2n+1}}\mF\la^{-2n-1}, \\
\bar{S}(\bla) &=& \sum_{n=0}^{\infty}\bt_{2n+1}\bla^{-2n-1}
          -\sum_{n=0}\frac{1}{2n+1}\pa_{\bt_{2n+1}}\mF\bla^{2n+1},
\eean
such that
\bea
\mB_{2n+1}(\la)&=&\left.\pa_{t_{2n+1}}S\right|_{\la\ \fixed}
          =\la^{2n+1}-\sum_{m=0}\frac{1}{2m+1}\mF_{2n+1,2m+1}\la^{-2m-1},
\label{B-1} \\
 \mB_{2n+1}(\bla)&=&\left.\pa_{t_{2n+1}}\bar{S}\right|_{\bla\ \fixed}
          =-\sum_{m=0}\frac{1}{2m+1}\mF_{2n+1,\wbar{2m+1}}\bla^{2m+1},
\label{B-2}\\
\bmB_{2n+1}(\la)&=&\left.\pa_{\bt_{2n+1}}S\right|_{\la\ \fixed}
          =-\sum_{m=0}\frac{1}{2m+1}\mF_{\wbar{2n+1},2m+1}\la^{-2m-1},
\label{bB-1}\\
          \bmB_{2n+1}(\bla)&=&\left.\pa_{\bt_{2n+1}}\bar{S}\right|_{\bla\ \fixed}
          =\bla^{-2n-1}-\sum_{m=0}\frac{1}{2m+1}
          \mF_{\wbar{2n+1},\wbar{2m+1}}\bla^{2m+1}.
\label{bB-2}
\eea
Noticing that (\ref{B-1}) and (\ref{bB-1}) are defined in $D_\infty$,
while (\ref{B-2}) and (\ref{bB-2}) in $D_0$.
From (\ref{B-1})-(\ref{bB-2}), for $n=0$ we have
\bea
p(\la)&=&\pa_{t_1}S(\la)=\la-\sum_{m=1}f_{2m}\la^{-2m+1}, \qquad
          f_{2m}=\frac{1}{2m-1}\mF_{1,2m-1},
\label{k(L)}\\
p(\bla)&=&\pa_{t_1}\bar{S}(\bla)=\sum_{m=0}\bar{f}_{2m}\bla^{2m+1}, \qquad
          \bar{f}_{2m}=-\frac{1}{2m+1}\mF_{1,\wbar{2m+1}},
\label{k(L)-2}\\
\bar{u}_0^{-1}p^{-1}(\la)&=&\pa_{\bt_1}S(\la)=\sum_{m=0}g_{2m}\la^{-2m-1}, \qquad
          g_{2m}=-\frac{1}{2m+1}\mF_{\wbar{1},2m+1},
\label{k(bL)-2} \\
\bar{u}_0^{-1}p^{-1}(\bla)&=&\pa_{\bt_1}\bar{S}(\bla)=\bla^{-1}-\sum_{m=1}\bar{g}_{2m}\bla^{2m-1}, \qquad
          \bar{g}_{2m}=\frac{1}{2m-1}\mF_{\wbar{1},\wbar{2m-1}}.
\label{k(bL)}
\eea
Therefore the dynamical variables $u_{2j}$ and $\bu_{2j}$ of the system are characterized by
the second derivatives of the free energy $\mF$.
In particular, from (\ref{k(L)}) and (\ref{k(bL)-2}), we have $u_2=\mF_{11}$
and $\bu_0^{-1}=-\mF_{1,\bar{1}}$, respectively.
%%%%%%%%%%%%%%%%%%%%%%%%%%%%%%%%%%%%%%%%%%%%%%%%%%%%%%%%%%%%
\subsection{Kernel formulas}
%%%%%%%%%%%%%%%%%%%%%%%%%%%%%%%%%%%%%%%%%%%%%%%%%%%%%%%%%%%%%
To employ kernel formula to derive dHirota equations of the EdBKP hierarchy
we shall first construct the generating functions of $\pa_pQ_{2n+1}(\la)$
and $\pa_p\bQ_{2n+1}(\bla)$. Let us follow the procedure for the derivation of dHirota
equations of the dToda case in the domains $D_\infty$, $D_0$, and $D_\infty \cap D_0$, separately.

In $D_\infty$. Multiplying  (\ref{k(L)}) by $\la^{2n-1}\pa_p\la$  for $n\geq 1$, we have
\[
 \la^{2n}\pa_p\la = p\la^{2n-1}\pa_p\la + \sum_{j=1}f_{2j}\la^{2n-2j}\pa_p\la,
\]
which, after taking the polynomial part, leads to the recurrence relation
\be
 \pa_pQ_{2n+1}(\la)=p(\la)\pa_pQ_{2n}(\la)+\sum_{j=1}^nf_{2j}\pa_pQ_{2n-2j+1}(\la),
\label{recur-1}
\ee
where $Q_{2n+1}\equiv\mB_{2n+1}/(2n+1)$ and $Q_{2n}\equiv(\la^{2n})_{\geq 0}/(2n)$.
Multiplying (\ref{recur-1}) by $\mu^{-2n}\ (n\geq 0)$ and summing over $n$ we obtain
\be
p(\mu)\sum_{n=0}\pa_pQ_{2n+1}(\la)\mu^{-2n-1}
 - p(\la)\sum_{n=0}\pa_pQ_{2n+2}(\la)\mu^{-2n-2} = 1.
\label{gen1}
\ee
On the other hand, multiplying both sides on (\ref{k(L)}) by $\la^{2n}\pa_p\la$ for $n\geq 1$,
it follows that
\[
 \la^{2n+1}\pa_p\la = p\la^{2n}\pa_p\la + \sum_{j=1}f_{2j}\la^{2n-2j+1}\pa_p\la.
\]
After taking the polynomial part we get another recurrence relation
\be
 \pa_pQ_{2n+2}(\la)=p(\la)\pa_pQ_{2n+1}(\la)+\sum_{j=1}^{n}f_{2j}\pa_pQ_{2n-2j+2}(\la).
\label{recur-2}
\ee
Multiplying (\ref{recur-2}) by $\mu^{-2n-1}\ (n\geq 0)$ and summing over $n$ we obtain
\be
p(\mu)\sum_{n=0}\pa_pQ_{2n+2}(\la)\mu^{-2n-2}
 = p(\la)\sum_{n=0}\pa_pQ_{2n+1}(\la)\mu^{-2n-1}.
\label{gen2}
\ee
Plugging (\ref{gen2}) into (\ref{gen1}) to eliminate $\sum_{n=0}\pa_pQ_{2n+2}(\la)\mu^{-2n-2}$,
we obtain the generating function of $\pa_pQ_{2n+1}(\la)$ as
\be
 \frac{p(\mu)}{p^2(\mu)-p^2(\la)}=\sum_{n=0}\pa_pQ_{2n+1}(\la)\mu^{-2n-1}.
\label{kernel}
\ee
Now we integrate the kernel (\ref{kernel}) with respect to $p(\la)$,
and normalize at $\mu=\infty$ to obtain
\be
 \frac{p(\mu)-p(\la)}{p(\mu)+p(\la)}
 =\exp\left(-2\sum_{n=0}^{\infty}Q_{2n+1}(\la)\mu^{-2n-1}\right).
\label{kQ}
\ee
If we denote $D(z)$ the differential operator
\[
D(z)=\sum_{n=0}^{\infty}\frac{z^{-2n-1}}{2n+1}\pa_{t_{2n+1}},
\]
then (\ref{kQ}) can be rewritten as
\be
 \frac{p(\mu)-p(\la)}{p(\mu)+p(\la)}=\exp\left(-2D(\mu)S(\la)\right)
 =\frac{\mu-\la}{\mu+\la}e^{2D(\la)D(\mu)\mF},
\label{dHirota-1}
\ee
which is the same result derived by Bogdanov and Konopelchenko \cite{BK} where
they obtained dHirota equations of the dBKP hierarchy from that of the dKP hierarchy
by taking into account the symmetry condition $S(-\la)=-S(\la)$.

In $D_0$. To obtain the generating function of $\pa_p\bQ_{2n+1}(\bla)$
we multiplying (\ref{k(bL)}) by $\bla^{-2n-1}\pa_p\bla$ for $n\geq 1$, then
\[
 \bla^{-2n-2}\pa_p\bla = \bar{u}_0^{-1}p^{-1}(\bla)\bla^{-2n-1}\pa_p\bla
                         + \sum_{j=1}\bar{g}_{2j}\bla^{-2n+2j-2}\pa_p\bla.
\]
Taking the negative part leads to the recurrence relation
\be
 \pa_p\bQ_{2n+1}(\bla)=\bar{u}_0^{-1}p^{-1}(\bla)\pa_p\bQ_{2n}(\bla)
                        +\sum_{j=1}^n\bar{g}_{2j}\pa_p\bQ_{2n-2j+1}(\bla),
\label{recur-b1}
\ee
where $\bQ_{2n+1}\equiv\bmB_{2n+1}/(2n+1)$ and $\bQ_{2n}\equiv(\bla^{-2n})_{\leq -1}/(2n)$.
Multiplying (\ref{recur-b1}) by $\bmu^{2n}\ (n\geq 0)$ and summing over $n$ we obtain
\be
 p^{-1}(\bmu)\sum_{n=0}\pa_p\bQ_{2n+1}(\bla)\bmu^{2n+1}
 - p^{-1}(\bla)\sum_{n=1}\pa_p\bQ_{2n}(\bla)\bmu^{2n} = -p^{-2}(\bla).
\label{genb1}
\ee
On the other hand, multiplying  (\ref{k(bL)}) by $\bla^{-2n}\pa_p\bla$ for $n\geq 1$,
it follows that
\[
 \bla^{-2n-1}\pa_p\bla = \bar{u}_0^{-1}p^{-1}(\bla)\bla^{-2n}\pa_p\bla
                         + \sum_{j=1}\bar{g}_{2j}\bla^{-2n+2j-1}\pa_p\bla.
\]
After taking the negative part we get another recurrence relation
\be
 \pa_p\bQ_{2n}(\bla)=\bar{u}_0^{-1}p^{-1}(\bla)\pa_p\bQ_{2n-1}(\bla)
                     +\sum_{j=1}^{n-1}\bar{g}_{2j}\pa_p\bQ_{2n-2j}(\bla).
\label{recur-b2}
\ee
Multiplying (\ref{recur-b2}) by $\bmu^{2n-1}\ (n\geq 1)$ and summing over $n$ we obtain
\be
 p^{-1}(\bmu)\sum_{n=1}\pa_p\bQ_{2n}(\bla)\bmu^{2n}
 = p^{-1}(\bla)\sum_{n=0}\pa_p\bQ_{2n+1}(\bla)\bmu^{2n+1}.
\label{genb2}
\ee
Plugging (\ref{genb2}) into (\ref{genb1}) to eliminate $\sum_{n=1}\pa_p\bQ_{2n}(\bla)\bmu^{2n}$,
which yields the generating function of $\pa_p\bQ_{2n+1}(\bla)$ as
\be
 \frac{p(\bmu)}{p^2(\bmu)-p^2(\bla)}=\sum_{n=0}\pa_p\bQ_{2n+1}(\bla)\bmu^{2n+1}.
\label{kernel-b}
\ee
Now we integrate the kernel (\ref{kernel-b}) with respect to $p(\bla)$,
and normalize at $\bmu=0$ to obtain
\be
 \frac{p(\bla)-p(\bmu)}{p(\bla)+p(\bmu)}
 =\exp\left(-2\sum_{n=0}^{\infty}\bQ_{2n+1}(\bla)\bmu^{2n+1}\right)
\label{bkQ}
\ee
which can be expressed as
\be
\frac{p(\bla)-p(\bmu)}{p(\bla)+p(\bmu)}
 =\frac{\bla-\bmu}{\bla+\bmu}e^{2\bD(\bla^{-1})\bD(\bmu^{-1})\mF}
\label{dHirota-2}
\ee
where
\[
\bD(z)=\sum_{n=0}^{\infty}\frac{z^{-2n-1}}{2n+1}\pa_{\bt_{2n+1}},
\]

In $D_\infty \cap D_0$. In view of (\ref{B-1}) and (\ref{B-2}),
 the functional $Q_{2n+1}(\la)$ and $p(\la)$ can be replaced by
$\pa_{t_{2n+1}}\bS(\bla)/(2n+1)$ and $\pa_{t_1}\bS(\bla)$, respectively.
Then (\ref{kQ}) becomes
\be
 \frac{p(\mu)-\pa_{t_1}\bS(\bla)}{p(\mu)+\pa_{t_1}\bS(\bla)}
 =e^{2D(\mu)\bD(\bla^{-1})\mF}.
\label{dHirota-3}
\ee
Similarly,  in view of (\ref{bB-1}) and (\ref{bB-2})
the functional $\bQ_{2n+1}(\bla)$ and $\bar{u}_0^{-1}p^{-1}(\bla)$
can be replaced by $\pa_{\bar{t}_{2n+1}}S(\la)/(2n+1)$ and $\pa_{\bar{t}_1}S(\la)$, respectively.
Then (\ref{bkQ}) becomes
\be
 \frac{\bar{u}_0^{-1}p^{-1}(\bmu)-\pa_{\bar{t}_1}S(\la)}
      {\bar{u}_0^{-1}p^{-1}(\bmu)+\pa_{\bar{t}_1}S(\la)}
 =e^{2D(\la)\bD(\bmu^{-1})\mF}.
\label{dHirota-4}
\ee
Note that, after making the replacements $\la\to\mu$ and $\bmu\to\bla$,
the right hand side of (\ref{dHirota-4}) coincides with that of (\ref{dHirota-3}). Hence
\be
 p(\mu)\pa_{\bar{t}_1}S(\mu) = \bar{u}_0^{-1}p^{-1}(\bla)\pa_{t_1}\bS(\bla)=\bar{u}_0^{-1}
 \label{extra}
\ee
where the last equality is determined by setting $\mu\rightarrow\infty$ and $\bla\rightarrow 0$:
\[
 \lim_{\mu\rightarrow\infty} p(\mu)\pa_{\bar{t}_1}S(\mu)
 =\lim_{\bla\rightarrow 0}\bar{u}_0^{-1}p^{-1}(\bla)\pa_{t_1}\bS(\bla)
 =-\mF_{1\bar{1}}=\bar{u}_0^{-1}.
\]
Eq. (\ref{extra}) contains the definitions (\ref{k(L)-2}) and (\ref{k(bL)-2})
and can be regarded as an extra condition besides the dHirota equations
(\ref{dHirota-1}), (\ref{dHirota-2}), and (\ref{dHirota-3}).

In summary, the EdBKP hierarchy can be characterized by a single function $\mF$
satisfying (\ref{dHirota-1}), (\ref{dHirota-2}), (\ref{dHirota-3}), and (\ref{dHirota-4})
(or  (\ref{dHirota-1}), (\ref{dHirota-2}), (\ref{dHirota-3}), and (\ref{extra}).).

%%%%%%%%%%%%%%%%%%%%%%%%%%%%%%%%%%%%%%%%%%%%%%%%%%%%%%%%%%
\subsection{Dispersionless Hirota equations}
%%%%%%%%%%%%%%%%%%%%%%%%%%%%%%%%%%%%%%%%%%%%%%%%%%%%%%%%%%%%%
Let us rewrite (\ref{dHirota-1}), (\ref{dHirota-2}), (\ref{dHirota-3}), and (\ref{dHirota-4})
as a set of equations satisfied by the second derivatives of $\mF$. \\
\textbf{(1)}\quad
We first rewrite the equation (\ref{dHirota-1}) as the form
\[
 \log\left(\frac{p(\la)-p(\mu)}{\la-\mu}\frac{\la+\mu}{p(\la)+p(\mu)}\right) = 2D(\la)D(\mu)\mF,
\]
where $p(\la)$ is defined by (\ref{k(L)}) and for $\la\rightarrow \mu$ we have
\bean
 \log(\mu\pa_{\mu}\log p(\mu))
&=& 2D^2(\mu)\mF, \\
&=& \sum_{j=1}^{\infty}\left(\sum_{\scriptstyle m+n=j-1 \atop\scriptstyle m,n\geq 0}
     \frac{2\mF_{2m+1,2n+1}}{(2m+1)(2n+1)}\right)\mu^{-2j}.
\eean
Then using the elementary Schur polynomial and (\ref{k(L)}) we obtain
\[
\frac{1+\sum_{m=0}^{\infty}\mF_{1,2m+1}\mu^{-2m-2}}{1-\sum_{m=0}^{\infty}\frac{1}{2m+1}\mF_{1,2m+1}\mu^{-2m-2}}
=\sum_{k=0}^{\infty}P_k(\theta_1,\theta_2,\ldots,\theta_k)\mu^{-2k},
\]
where $\theta_i$, $i\geq 1$ are defined by
\[
 \theta_i=\sum_{\scriptstyle m+n=i-1 \atop\scriptstyle m,n\geq 0}\frac{2\mF_{2m+1,2n+1}}{(2m+1)(2n+1)}.
\]
By comparing the coefficients of $\mu^{-2k}$ on both sides we obtain a set
of relations satisfied by $\mF_{2i+1,2j+1}$:
\be
 \frac{2m+2}{2m+1}\mF_{1,2m+1}-P_{m+1}(\theta)
 +\sum_{\scriptstyle j+k=m-1 \atop\scriptstyle j,k\geq 0}
 \frac{1}{2j+1}\mF_{1,2j+1}P_{k+1}(\theta)=0.
 \label{Fmn-1}
\ee
\textbf{(2)}\quad
Similarly, we consider (\ref{dHirota-2}) by taking the limit $\bla\rightarrow \bmu$. Then
\bean
 \log(\bmu\pa_{\bmu}\log p(\bmu))
&=& 2D^2(\bmu)\mF, \\
&=& \sum_{j=1}^{\infty}\left(\sum_{\scriptstyle m+n=j-1 \atop\scriptstyle m,n\geq 0}
     \frac{2\mF_{\wbar{2m+1},\wbar{2n+1}}}{(2m+1)(2n+1)}\right)\bmu^{2j}.
\eean
Also, using the elementary Schur polynomial and (\ref{k(bL)}) we obtain
\[
\frac{1+\sum_{m=0}^{\infty}\mF_{\bar{1},\wbar{2m+1}}\bmu^{2m+2}}
     {1-\sum_{m=0}^{\infty}\frac{1}{2m+1}\mF_{\bar{1},\wbar{2m+1}}\bmu^{2m+2}}
=\sum_{k=0}^{\infty}P_k(\bth_1,\bth_2,\ldots,\bth_k)\bmu^{2k},
\]
where $\bth_i$, $i\geq 1$ are defined by
\[
 \bth_i=\sum_{\scriptstyle m+n=i-1 \atop\scriptstyle m,n\geq 0}
        \frac{2\mF_{\wbar{2m+1},\wbar{2n+1}}}{(2m+1)(2n+1)}.
\]
By comparing the coefficients of $\bmu^{2k}$ on both sides we obtain a set
of relations satisfied by $\mF_{\wbar{2i+1},\wbar{2j+1}}$:
\be
\frac{2m+2}{2m+1}\mF_{\bar{1},\wbar{2m+1}}-P_{m+1}(\bth)
 +\sum_{\scriptstyle j+k=m-1 \atop\scriptstyle j,k\geq 0}
   \frac{1}{2j+1}\mF_{\bar{1},\wbar{2j+1}}P_{k+1}(\bth)=0.
   \label{Fmn-2}
\ee
\textbf{(3)}\quad
From (\ref{dHirota-3}), by taking $\bla\rightarrow 1/\mu$ and using (\ref{k(L)})
and (\ref{k(L)-2}), it is straightforward to obtain
\[
 \frac{1-\sum_{m=0}^{\infty}\frac{1}{2m+1}(\mF_{1,2m+1}-\mF_{1,\wbar{2m+1}})\mu^{-2m-2}}
 {1-\sum_{m=0}^{\infty}\frac{1}{2m+1}(\mF_{1,2m+1}+\mF_{1,\wbar{2m+1}})\mu^{-2m-2}}
= \sum_{k=0}^{\infty}P_k(\tth_1,\tth_2,\ldots,\tth_k)\mu^{-2k},
\]
where $P_k(\tth)$ are  elementary Schur polynomials and $\tth_i$, $i\geq 1$ are defined by
\[
 \tth_i = \sum_{\scriptstyle m+n=i-1 \atop \scriptstyle m,n\geq 0}
          \frac{2\mF_{2n+1,\wbar{2m+1}}}{(2m+1)(2n+1)}.
\]
By comparing the coefficients of $\mu^{-2k}$ on both sides we obtain a set
of relations satisfied by $\mF_{2i+1,\wbar{2j+1}}$:
\be
 \frac{2}{2m+1}\mF_{1,\wbar{2m+1}}-P_{m+1}(\tth)
 +\sum_{\scriptstyle j+k=m-1 \atop \scriptstyle j,k\geq 0}
 \frac{1}{2j+1}(\mF_{1,2j+1}+\mF_{1,\wbar{2j+1}})P_{k+1}(\tth) = 0.
 \label{Fmn-3}
\ee
\textbf{(4)}\quad
From (\ref{dHirota-4}) it is straightforward to taking $\bmu\rightarrow 1/\la$ and  by
(\ref{k(bL)}) and (\ref{k(bL)-2}) we have
\[
 \frac{1-\sum_{m=0}^{\infty}\frac{1}{2m+1}(\mF_{\bar{1},\wbar{2m+1}}-\mF_{\bar{1},2m+1})\la^{-2m-2}}
 {1-\sum_{m=0}^{\infty}\frac{1}{2m+1}(\mF_{\bar{1},\wbar{2m+1}}+\mF_{\bar{1},2m+1})\la^{-2m-2}}
= \sum_{k=0}^{\infty}P_k(\tth_1,\tth_2,\ldots,\tth_k)\la^{-2k},
\]
where $P_k(\tth)$ is defined as before and the above equation can be simplified as
\be
 \frac{2}{2m+1}\mF_{\bar{1},2m+1}-P_{m+1}(\tth)
 +\sum_{\scriptstyle j+k=m-1 \atop \scriptstyle j,k\geq 0}
 \frac{1}{2j+1}(\mF_{\bar{1},\wbar{2j+1}}+\mF_{\bar{1},2j+1})P_{k+1}(\tth) = 0.
 \label{Fmn-4}
\ee
Note that for $u_2=\mF_{11}$ the first equation in (\ref{Fmn-1}) is
nothing but the 2+1-dimensional dBKP equation\cite{DR92}
\[
 3u_t+15u^2u_x-5uu_y-5u_x\pa_x^{-1}u_y-\frac{5}{3}\pa_x^{-1}u_{yy}=0,
\]
where $t_1=x$, $t_3=y$, $t_5=t$ and $u\equiv u_2$, while
the first equation in (\ref{Fmn-3}) is the simplest nontrivial 2+1-dimensional equation
of EdBKP hierarchy involving $t_1=x$, $t_3=y$, and $\bar{t}_1$:
\[
 uu_{\bar{t}_1} + u_x\pa_x^{-1}u_{\bar{t}_1} - \frac{1}{3}\pa_x^{-1}u_{y\bar{t}_1}=0.
\]
For convenience, we list other higher flows of (\ref{Fmn-1})-(\ref{Fmn-4}) in appendix B.

%%%%%%%%%%%%%%%%%%%%%%%%%%%%%%%%%%%%%%%%%%%%%%%%%%%%%%%%%%%%%%%%%%%
\section{Associativity equations in EdBKP hierarchy}
%%%%%%%%%%%%%%%%%%%%%%%%%%%%%%%%%%%%%%%%%%%%%%%%%%%%%%%%%%%%%%%%%%%%
\subsection{WDVV equations and generating functions}
%%%%%%%%%%%%%%%%%%%%%%%%%%%%%%%%%%%%%%%%%%%%%%%%%%%%%%%%%%%%%%%%%
Following \cite{BMRWZ}, let us briefly recall the basic notions of associativity equations.
Let $\mF$ be a differentiable functions of a set of time variables $t=(t_1,t_2,\cdots)$; namely,
$ \mF=\mF(t_1,t_2,t_3,\ldots)$.
Then define $\mF_{ij}=\pa_{t_i}\pa_{t_j}\mF$, $\mF_{ijk}=\pa_{t_i}\pa_{t_j}\pa_{t_k}\mF$.
One can choose one of the time variable, say $t_1$, and assume the nondegenerate metric
$\eta_{ij}=\mF_{ij1}$, which can be regarded as a transform from $\{t_i\}$ to $\{\mF_{j1}\}$.
Define the matrix $C^l_{ij}=\pa\mF_{ij}/\pa\mF_{l1}$, then
\be
 \mF_{ijk} = \sum_lC^l_{ij}\eta_{lk}=\sum_lC^l_{ij}\mF_{lk1}
 \label{def-c}
\ee
connects $\mF_{ijk}$ and $\eta_{ij}$.
It can also be treated as a definition of $C^l_{ij}$.

If $C^l_{ij}$ is the structure constant of an associativity algebra generated by $\{\phi_i\}$;
namely, $\phi_i\cdot\phi_j=\sum_lC^l_{ij}\phi_l$, then we have the constraints
$\sum_lC^l_{ij}C^n_{lk}=\sum_lC^l_{jk}C^n_{il}$, or equivalently
\[
 \sum_lC^l_{ij}\mF_{lkm} = \sum_lC^l_{jk}\mF_{ilm},
\]
which is the so-called WDVV equations
arising from topological field theories \cite{W,DVV91}.
Thus
\be
 X_{ijkn}\equiv\sum_lC^l_{ij}\mF_{lkn}
\label{X}
\ee
is symmetric with respect to permutations of any indices.
In general, one may choose any $t_a$ to define the metric $\eta(a)_{ij}=\mF_{ija}$
and the structure constants $C^l_{ij}(a)=\pa\mF_{ij}/\pa\mF_{la}$, then
they obey the same associativity relations
\be
 \sum_lC^l_{ij}(a)\mF_{lkm} = \sum_lC^l_{jk}(a)\mF_{ilm}.
 \label{wdvv}
\ee
Introducing matrices $\mathbf{F}_a$ with matrix elements $(\mathbf{F}_a)_{ij}=\mF_{ija}$,
one can rewrite the WDVV equations as
\[
 \mathbf{F}_j\mathbf{F}_a^{-1}\mathbf{F}_m = \mathbf{F}_m\mathbf{F}_a^{-1}\mathbf{F}_j.
\]
In the case of infinitely many variables it is convenient to define generating functions
for $\mF_{ij}$, $\mF_{ijk}$, $C_{ij}$ and $X_{ijkl}$ as follows
\bean
D_1D_2\mF&=&\sum_{i,j=1}^\infty\frac{z_1^{-i}}{i}\frac{z_2^{-j}}{j}\mF_{ij},\\
D_1D_2D_3\mF&=&\sum_{i,j,k=1}^\infty\frac{z_1^{-i}}{i}\frac{z_2^{-j}}{j}\frac{z_3^{-k}}{k}\mF_{ijk},\\
C^l(z_1,z_2)&=&\sum_{i,j=1}^\infty \frac{z_1^{-i}}{i}\frac{z_2^{-j}}{j}C_{ij}^l,\\
X(z_1,z_2,z_3,z_4)&=&\sum_{i,j,k,n=1}^\infty\frac{z_1^{-i}}{i}\frac{z_2^{-j}}{j}\frac{z_3^{-k}}{k}
\frac{z_3^{-n}}{n}X_{ijkn},
\eean
where $D_i\equiv D(z_i)=\sum_{j=1}^\infty z_i^{-j}\pa_{t_j}/j$.
From the definition (\ref{X}) it is easy to see that $X_{ijkn}$ is symmetric in
$(ij)$ and $(kn)$, while the associativity equations (\ref{wdvv}) implies that
$X_{ijkn}=X_{ikjn}$ or, in terms of generating function,
\be
X(z_1,z_2,z_3,z_4)=X(z_1,z_3,z_2,z_4).
\label{permX}
\ee
In \cite{BMRWZ}, the associativity equations of the dKP and dToda hierarchies were derived
from the corresponding dHirota equations by verifying the symmetric property (\ref{permX}).

%%%%%%%%%%%%%%%%%%%%%%%%%%%%%%%%%%%%%%%%%%%%%%%%%%%%%%%%%%%%%%%%%%
\subsection{From dHirota to WDVV}
%%%%%%%%%%%%%%%%%%%%%%%%%%%%%%%%%%%%%%%%%%%%%%%%%%%%%%%%%%%%%%%%%%%
To derive WDVV equations from dHirota equations for EdBKP we set $\bar{t}_{-2k-1}=t_{2k+1}$
($k\leq -1$) for convenience and rewrite the operator $\bD(z)$ as
\[
 \bD(z)=\sum_{k\leq -1}\frac{-z^{2k+1}}{2k+1}\pa_{t_{2k+1}}.
\]
We will use the notation $D(z_i)=D_i$ and $\bD(z_i)=\bD_i$ for brevity.
The functions
\[
 D_1D_2\mF=\sum_{k,m\geq 0}\frac{z_1^{-2k-1}z_2^{-2m-1}}{(2k+1)(2m+1)}\mF_{2k+1,2m+1}
\]
and $D_1D_2D_3\mF$ generate separately the sets of $\mF_{2k+1,2m+1}$ and
$\mF_{2k+1,2m+1,2n+1}$ for all $k,m,n\geq 0$. On the other hand, the functions
\[
 D_1\bD_2\mF=\sum_{k\geq 0,m\leq -1}
                   \frac{-z_1^{-2k-1}z_2^{2m+1}}{(2k+1)(2m+1)}\mF_{2k+1,2m+1}
\]
generate the set of $\mF_{2k+1,2m+1}$ for all $k\geq 0$, and $m\leq -1$, etc.
We also introduce generating functions for the structure constants $C^{2l+1}_{2i+1,2j+1}$:
\bea
 C^l(z_1^+,z_2^+)&=&\sum_{i,j\geq 0}C^{2l+1}_{2i+1,2j+1}\frac{z_1^{-2i-1}z_2^{-2j-1}}{(2i+1)(2j+1)}, \no\\
 C^l(z_1^-,z_2^-)&=&\sum_{i,j\leq -1}C^{2l+1}_{2i+1,2j+1}\frac{z_1^{2i+1}z_2^{2j+1}}{(2i+1)(2j+1)}, \no\\
 C^l(z_1^+,z_2^-)&=&\sum_{i\geq 0,j\leq -1}C^{2l+1}_{2i+1,2j+1}\frac{-z_1^{-2i-1}z_2^{2j+1}}{(2i+1)(2j+1)},
 \label{gen-str}
\eea
and the $X_{ijkn}$ :
\be
 X(z_1,z_2,z_3,z_4)\equiv \sum_{i,j,k,n=-\infty}^{\infty}
 \frac{z_1^{-|i|}}{|i|} \frac{z_2^{-|j|}}{|j|}
 \frac{z_3^{-|k|}}{|k|} \frac{z_4^{-|n|}}{|n|}  X_{ijkn},
 \qquad i,j,k,n\in \mathbf{Z_{odd}}.
\label{X1234}
\ee
Notice that from (\ref{gen-str}) the property holds: $C^l(z_1^+,z_2^-)=C^l(z_2^-,z_1^+)$.
The infinite WDVV equations are thus equivalent to the symmetry of the $X(z_1,z_2,z_3,z_4)$
under permutations of $z_1,z_2,z_3,z_4$.
In the following, we shall show that  the associativity equations in the EdBKP hierarchy
can be verified in a similar way as the dToda case \cite{BMRWZ}.

Let us rewrite the dHirota equations (\ref{dHirota-1}), (\ref{dHirota-2}), and (\ref{dHirota-3})
 as
\bea
 \frac{p_1-p_2}{p_1+p_2} &=& \frac{z_1-z_2}{z_1+z_2}e^{2D_1D_2\mF},
\label{T-I} \\
 \frac{\bp_1-\bp_2}{\bp_1+\bp_2} &=& \frac{z_1-z_2}{z_1+z_2}e^{2\bD_1\bD_2\mF},
\label{T-II} \\
 \frac{p_1+\bD_2\pa_{t_1}\mF}{p_1-\bD_2\pa_{t_1}\mF} &=& e^{2D_1\bD_2\mF},
\label{T-III}
\eea
where $p_i\equiv p(z_i)=z_i-D_i\pa_{t_1}\mF$ and
$\bp_i\equiv \bu_0^{-1}p^{-1}(z_i^{-1})=z_i-\bD_i\pa_{t_{-1}}\mF$.
The symmetric form of equations (\ref{T-I}) and (\ref{T-II}) are\cite{BK}
\bea
 f_{12}f_{23}f_{31} + f_{12}+f_{23}+f_{31} &=& 0,
\label{sym-1} \\
 \brf_{12}\brf_{23}\brf_{31} + \brf_{12}+\brf_{23}+\brf_{31} &=& 0,
\label{sym-2}
\eea
where
\bean
 f_{ij} &=& \frac{p_i-p_j}{p_i+p_j}
         = \frac{z_i-z_j}{z_i+z_j}e^{2D_iD_j\mF}, \quad 1\leq i,j \leq 3,  \\
 \brf_{ij} &=& \frac{\bp_i-\bp_j}{\bp_i+\bp_j}
            = \frac{z_i-z_j}{z_i+z_j}e^{2\bD_i\bD_j\mF}, \quad 1\leq i,j \leq 3,
\eean
By taking the $D_4$ and $\bD_4$-derivatives of (\ref{sym-1}) we get the following
useful identities
\bea
 p_1^2(p_3D_2-p_2D_3)D_1D_4\mF + p_2^2(p_1D_3-p_3D_1)D_2D_4\mF + p_3^2(p_2D_1-p_1D_2)D_3D_4\mF =0,
\label{idn-1} \\
 p_1^2(p_3D_2-p_2D_3)D_1\bD_4\mF + p_2^2(p_1D_3-p_3D_1)D_2\bD_4\mF + p_3^2(p_2D_1-p_1D_2)D_3\bD_4\mF =0.
\label{idn-2}
\eea
Similarly, we have another two identities if we apply $D_4$ and $\bD_4$ to (\ref{sym-2}).
We define a non-degenerate metric to be $\eta_{2i+1,2j+1}=\mF_{2i+1,2j+1,1}$,
 and the structure constants as
\[
\mF_{2i+1,2j+1,2k+1} = \sum_{l=-\infty}^{\infty}C^{2l+1}_{2i+1,2j+1}\mF_{2l+1,2k+1,1}.
\label{cij}
\]
where the indices take integer values.
From the definition of the metric, we have $C^{2l+1}_{1,2j+1}=\delta^l_j$ for all $l,j$.
To find other structure constants, apply $\pa_{t_{2k+1}}$ to the dHirota equations
(\ref{T-I})-(\ref{T-III}):
\bea
 D_1D_2\pa_{2k+1}\mF &=& \frac{p_1D_2-p_2D_1}{p_1^2-p_2^2}\mF_{2k+1,1},
\label{k1} \\
 \bD_1\bD_2\pa_{2k+1}\mF &=& \frac{\bp_1\bD_2-\bp_2\bD_1}{\bp_1^2-\bp_2^2}\mF_{2k+1,-1},
\label{k2} \\
 D_1\bD_2\pa_{2k+1}\mF &=& \frac{p_1\bD_2+(\bD_2\pa_{t_1}\mF)D_1}{p_1^2-(\bD_2\pa_{t_1}\mF)^2}\mF_{2k+1,1},
\label{k3}
\eea
Comparing with  (\ref{def-c}) the generating functions of structure constants (\ref{gen-str})
can be read out from the right-hand side of these equations.
However, the r.h.s. of equation (\ref{k2}) is related to the
derivation of $\mF$ with respect to $t_{-1}$, not $t_1$, which forbids us
to extract  structure constants from (\ref{k2}) directly.
Fortunately, this problem can be removed by considering (\ref{extra}) in the form
\be
D_1\pa_{t_{-1}}\mF=\frac{\bp_2}{p_1}\bD_2\pa_{t_1}\mF.
\label{relation}
\ee
Differentiating above with respect to $t_{2k+1}$ yields
\[
D_1\mF_{2k+1,-1}=\frac{\pa_{2k+1}\bp_2}{p_1}\bD_2\pa_{t_1}\mF+\frac{\bp_2}{p_1}\bD_2\mF_{2k+1,1}-
\frac{\bp_2\pa_{2k+1}p_1}{p_1^2}\bD_2\pa_{t_1}\mF
\]
which together with $\pa_{2k+1}p_1=-D_1\mF_{2k+1,1}$ and
$\pa_{2k+1}\bp_2=-\bD_2\mF_{2k+1,-1}$ implies that
\be
D_1\mF_{2k+1,-1}=-\frac{\bD_2\pa_{t_1}\mF}{p_1}\bD_2\mF_{2k+1,-1}+\frac{\bp_2}{p_1}\bD_2\mF_{2k+1,1}
+\frac{\bp_2}{p_1^2}(\bD_2\pa_{t_1}\mF)D_1\mF_{2k+1,1}.
\label{relation-1}
\ee
Taking the limit $z_1\to \infty$ in (\ref{relation-1}) we obtain
\be
\bD_2\mF_{2k+1,-1}=\frac{\bp_2}{\bD_2\mF_1}\bD_2\mF_{2k+1,1},
\label{relation-2}
\ee
which, after substituting back into (\ref{relation-1}), yields
\[
D_1\mF_{2k+1,-1}=\frac{\bp_2}{p_1^2}(\bD_2\pa_{t_1}\mF)D_1\mF_{2k+1,1}
\]
On the other hand, exchanging the variables $z_1\leftrightarrow z_2$, we have
\bea
 \bD_1\mF_{2k+1,-1} &=& \frac{\bp_1}{\bD_1\mF_1}\bD_1\mF_{2k+1,1},
\label{relation-3}  \\
 D_2\mF_{2k+1,-1} &=& \frac{\bp_1}{p_2^2}(\bD_1\mF_1)D_2\mF_{2k+1,1}. \no
\eea
Substituting (\ref{relation-2}) and (\ref{relation-3}) into the numerator of (\ref{k2})
and taking into account (\ref{relation}), then  (\ref{k2}) becomes
\be
 \bD_1\bD_2\pa_{2k+1}\mF =
 \frac{(\bD_2\pa_{t_1}\mF)\bD_1-(\bD_1\pa_{t_1}\mF)\bD_2}{(\bD_1\pa_{t_1}\mF)^2-(\bD_2\pa_{t_1}\mF)^2}\mF_{2k+1,1}.
\label{k2-2}
\ee
Now we can find all structure constants by using equations (\ref{k1}), (\ref{k3}),
and (\ref{k2-2}).
First, we conclude from (\ref{k1}) and (\ref{k2-2}) that $C^{2l+1}_{2i+1,2j+1}=0$ for
$i,j\geq 0$ and $l\leq -1$, or $i,j\leq -1$ and $l\geq 0$.
Next, if all the indices are positive, we have:
\be
C^l(z_1^+,z_2^+)
= \frac{p_1z_2^{-2l-1}-p_2z_1^{-2l-1}}{(2l+1)(p_1^2-p_2^2)}, \quad l\geq 0,
\label{str1}
\ee
while all the indices are negative, it gives
\be
C^l(z_1^-,z_2^-)
=\frac{(\bD_1\pa_{t_1}\mF)z_2^{2l+1}-(\bD_2\pa_{t_1}\mF)z_1^{2l+1}}{(2l+1)\left((\bD_1\pa_{t_1}\mF)^2-(\bD_2\pa_{t_1}\mF)^2\right)},
 \quad l\leq -1.
\label{str2}
\ee
When $i$ and $j$ have different signs we use  (\ref{k3})  to obtain:
\be
C^l(z_1^+,z_2^-)= \left\{\ba{ll}
\displaystyle
    \frac{(\bD_2\pa_{t_1}\mF)z_1^{-2l-1}}{(2l+1)(p_1^2-(\bD_2\pa_{t_1}\mF)^2)}, & l\geq 0, \\
\displaystyle
    \frac{-p_1z_2^{2l+1}}{(2l+1)(p_1^2-(\bD_2\pa_{t_1}\mF)^2)}, & l\leq -1.
    \ea\right.
\label{str3}
\ee

Making use of the structure constants, we show that any solution $\mF$ of
the EdBKP hierarchy obeys the WDVV equations:
\be
 \sum_{l=-\infty}^{\infty}C^{2l+1}_{2i+1,2j+1}\mF_{2l+1,2k+1,2n+1}
= \sum_{l=-\infty}^{\infty}C^{2l+1}_{2i+1,2k+1}\mF_{2l+1,2j+1,2n+1},
  \quad i,j,k,n \in \mathbf{Z}.
\label{WDVV}
\ee
Although the generating functions (\ref{X1234}) is totally symmetric w.r.t. permutations of
$z_1,\ldots,z_4$, however, it is enough to prove the symmetry w.r.t. the permutations
of $z_2$ and $z_3$ in (\ref{X1234}), i.e. $X(z_1,z_2,z_3,z_4)=X(z_1,z_3,z_2,z_4)$.

For all positive indices in (\ref{WDVV}), we use (\ref{str1}) to obtain the generating function
\[
 X(z_1,z_2,z_3,z_4) = \frac{1}{p_1^2-p_2^2}(p_1D_2-p_2D_1)D_3D_4\mF.
\label{X-I}
\]
Hence we have to prove
\be
 (p_1^2-p_3^2)(p_1D_2-p_2D_1)D_3D_4\mF = (p_1^2-p_2^2)(p_1D_3-p_3D_1)D_2D_4\mF.
\label{X23-1}
\ee
It is straightforward to bring (\ref{X23-1}) into the identity (\ref{idn-1}) by
eliminating $p_1^3D_2D_3D_4\mF$ to the both sides of (\ref{X23-1}),
which concludes the proof for this case. If the index $n$ in (\ref{WDVV}) is negative,
then with the help of (\ref{idn-2}) the same arguments go through.

Let all the indices in (\ref{WDVV}) be negative, then by (\ref{str2}) we have
the following equation to be verified:
\bean
0 &=& (\bD_1\pa_{t_1}\mF)^2\left((\bD_3\pa_{t_1}\mF)\bD_2-(\bD_2\pa_{t_1}\mF)\bD_3\right)\bD_1\bD_4\mF \\
   && + (\bD_2\pa_{t_1}\mF)^2\left((\bD_1\pa_{t_1}\mF)\bD_3-(\bD_3\pa_{t_1}\mF)\bD_1\right)\bD_2\bD_4\mF \\
   && + (\bD_3\pa_{t_1}\mF)^2\left((\bD_2\pa_{t_1}\mF)\bD_1-(\bD_1\pa_{t_1}\mF)\bD_2\right)\bD_3\bD_4\mF,
\eean
which, after consulting (\ref{k2-2}), is indeed an identity.
This completes the proof for this case.

Let $j$ in (\ref{WDVV}) be negative, and all others be positive.
In terms of the generating functions the WDVV equations now reads
\be
 \frac{(p_1\bD_2+(\bD_2\pa_{t_1}\mF)D_1)D_3D_4\mF}{p_1^2-(\bD_2\pa_{t_1}\mF)^2}
 = \frac{(p_1D_3-p_3D_1)\bD_2D_4\mF}{p_1^2-p_3^2}.
\label{X23-3}
\ee
We express $D_3\bD_2D_4$ and $D_1\bD_2D_4$ from (\ref{k3})
(multiplying it by $z_4^{-2k-1}/(2k+1)$, and summing over $k$) and
substitute them back into (\ref{X23-3}).
It follows that
\[
 D_1D_3D_4\mF= \frac{(p_1D_3-p_3D_1)D_4\pa_{t_1}\mF}{p_1^2-p_3^2},
\]
which is provided by (\ref{k1}). We conclude the proof of this case.

Finally, let $j,k$ be negative, and $i,n$ be positive. In this case we have the following
WDVV equations to prove:
\be
 \frac{\left(p_1\bD_2+(\bD_2\pa_{t_1}\mF)D_1\right)\bD_3D_4\mF}{p_1^2-(\bD_2\pa_{t_1}\mF)^2}
= \frac{\left(p_1\bD_3+(\bD_3\pa_{t_1}\mF)D_1\right)\bD_2D_4\mF}{p_1^2-(\bD_3\pa_{t_1}\mF)^2},
\label{X23-4}
\ee
where we have used $C^l(z_2^-,z_1^+)$ and $C^l(z_3^-,z_1^+)$ defined in (\ref{str3})
to the l.h.s and r.h.s. of (\ref{X23-4}), respectively.
Expressing $D_1\bD_3D_4$ and $D_1\bD_2D_4$ from (\ref{k3})
and substituting them back into (\ref{X23-4}), then
\[
 \bD_2\bD_3D_4\mF
 = \frac{\left((\bD_3\pa_{t_1}\mF)\bD_2-(\bD_2\pa_{t_1}\mF)\bD_3\right)D_4\pa_{t_1}\mF}{(\bD_2\pa_{t_1}\mF)^2-(\bD_3\pa_{t_1}\mF)^2}
\]
which is provided by (\ref{k2-2}).
This completes the proof of WDVV equations for the EdBKP hierarchy.
%%%%%%%%%%%%%%%%%%%%%%%%%%%%%%%%%%%%%%%%%%%%%%%%%%%%%%%%%%%%%%%%%%%%%%%%%
\subsection{Associative algebra and residue formula}
%%%%%%%%%%%%%%%%%%%%%%%%%%%%%%%%%%%%%%%%%%%%%%%%%%%%%%%%%%%%%%%%%%%%%%%%%
The realization of the associative algebra with the structure constants (\ref{str1})--(\ref{str3}),
is obtained with the help of the kernel formulas (\ref{kernel}) and (\ref{kernel-b})
by introducing the generators
\be
 \phi_{2k+1}(p) =\frac{d\mB_{2k+1}(p)}{dp}, \quad k\in\mathbf{Z}
\ee
where $\mB_{2k+1}\equiv \bmB_{-2k-1}$ for $k\leq -1$.
Then the kernel formulas (\ref{kernel}) and (\ref{kernel-b}) give us the following relations:
\bea
 \frac{p(z)}{p^2-p^2(z)} &=& -\sum_{k\geq 0}\frac{z^{-2k-1}}{2k+1}\phi_{2k+1}(p),
\label{re-1}  \\
 \frac{\bD(z)\pa_{t_1}\mF}{p^2-(\bD(z)\pa_{t_1}\mF)^2} &=& -\sum_{k\leq -1}\frac{z^{2k+1}}{2k+1}\phi_{2k+1}(p).
\label{re-2}
\eea
For example, we start by writing the identity
\be
  \frac{p_1p_2}{(p^2-p_1^2)(p^2-p_2^2)}
= \frac{p_1p_2}{p_1^2-p_2^2}\left(\frac{1}{p^2-p_1^2}-\frac{1}{p^2-p_2^2}\right),
\label{id-1}
\ee
and expanding both sides in $z_1^{-1}, z_2^{-1}$.
Then, using (\ref{re-1}) and (\ref{id-1}), and comparing with (\ref{str1}), we obtain the algebra
\[
 \phi_{2i+1}(p)\phi_{2j+1}(p)=\sum_{l\geq 0}C^{2l+1}_{2i+1,2j+1}\phi_{2l+1}(p), \quad i,j\geq 0.
\]
Next, we can expand both sides of the identity
\be
 \frac{p_1\bD_2\pa_{t_1}\mF}{(p^2-p_1^2)(p^2-(\bD_2\pa_{t_1}\mF)^2)}
= \frac{p_1\bD_2\pa_{t_1}\mF}{p_1^2-(\bD_2\pa_{t_1}\mF)^2}
  \left(\frac{1}{p^2-p_1^2}-\frac{1}{p^2-(\bD_2\pa_{t_1}\mF)^2}\right)
  \label{id-2}
\ee
in $z_1^{-1},z_2$. Using (\ref{re-1}), (\ref{re-2}) and (\ref{id-2}), and comparing with
 (\ref{str3}), we have the algebra
\[
 \phi_{2i+1}(p)\phi_{2j+1}(p)=\sum_{l=-\infty}^{\infty}C^{2l+1}_{2i+1,2j+1}\phi_{2l+1}(p),
 \quad i\geq 0,\; j\leq -1.
\]
The other algebra for $i,j,l\leq -1$ can be verified in a similar way and the structure
constants are given by (\ref{str2}).

Now let us derive the residue formulas for third order derivatives of $\mF$
(i.e. $\mF_{2j+1,2k+1,2m+1}$) directly from dHirota equations.
For the case of all the indices are positive, using (\ref{k1}) for $k=0$
\[
 D_1D_2\pa_{t_1}\mF = \frac{p_1D_2-p_2D_1}{p_1^2-p_2^2}\pa_{t_1}^2\mF,
\]
we can express $D_1D_2D_3\mF$  in terms of $D_i\pa_{t_1}^2\mF$ only:
\be
D_1D_2D_3\mF = 2\sum_{i=1}^3\res_{p_i}
               \left(
               \frac{p_1p_2p_3D(z(p))\pa_{t_1}^2\mF}{(p^2-p_1^2)(p^2-p_2^2)(p^2-p_3^2)}dp
               \right),
\label{res:+++}
\ee
Expanding both sides of (\ref{res:+++}) in the series in $z_1,z_2,z_3$ and using (\ref{re-1})
we obtain
\[
 \mF_{2j+1,2k+1,2m+1} =
 \frac{-1}{\pi i}\oint_{C_{\infty}}\frac{\pa_{t_1}z(p)}{z'(p)}
 \phi_{2j+1}(p)\phi_{2k+1}(p)\phi_{2m+1}(p)dp,   \quad j,k,m \geq 0.
\]
where, due to the fact that $0 =\pa_{t_1}p(p)=\pa_{t_1}p(z)+\pa_z p(z)\pa_{t_1}z$,
we have rewrited $D(z(p))\pa_{t_1}^2\mF=-\pa_{t_1}p(z)=\pa_{t_1}z/z'(p)$ in the numerator of (\ref{res:+++}).

Similarly, for those cases containing non-positive indices we have
\bean
D_1D_2\bD_3\mF &=& 2\sum_{i=1}^3\res_{p_i}
               \left(
               \frac{p_1p_2(\bD_3\pa_{t_1}\mF)\pa_{t_1}p}{(p^2-p_1^2)(p^2-p_2^2)(p^2-(\bD_3\pa_{t_1}\mF)^2)}
               dp
               \right), \\
D_1\bD_2\bD_3\mF &=& 2\sum_{i=1}^3\res_{p_i}
               \left(
               \frac{-p_1(\bD_2\pa_{t_1}\mF)(\bD_3\pa_{t_1}\mF)\pa_{t_1}p}
                    {(p^2-p_1^2)(p^2-(\bD_2\pa_{t_1}\mF)^2)(p^2-(\bD_3\pa_{t_1}\mF)^2)}dp
               \right),  \\
\bD_1\bD_2\bD_3\mF &=& 2\sum_{i=1}^3\res_{p_i}
               \left(
               \frac{(\bD_1\pa_{t_1}\mF)(\bD_2\pa_{t_1}\mF)(\bD_3\pa_{t_1}\mF)\pa_{t_1}p}
                    {(p^2-(\bD_1\pa_{t_1}\mF)^2)(p^2-(\bD_2\pa_{t_1}\mF)^2)(p^2-(\bD_3\pa_{t_1}\mF)^2)}
                    dp
               \right),
\eean
where $\pa_{t_1}p_i=-D_i\pa_{t_1}^2\mF$ when we evaluate the residue at $p=p_i$, while
 $\pa_{t_1}p_i=\bD_i\pa_{t_1}^2\mF$ at $p=\bD_i\pa_{t_1}\mF$.
Using (\ref{re-1}) and (\ref{re-2}), similar residue formulas can be written down.

%%%%%%%%%%%%%%%%%%%%%%%%%%%%%%%%%%%%%%%%%%%%%%%%%%%%%%%%%%%
\section{Concluding remarks}
%%%%%%%%%%%%%%%%%%%%%%%%%%%%%%%%%%%%%%%%%%%%%%%%%%%%%%%%%%%%
We have demonstrated the method of kernel formula to derive dHirota equations
for several dispersionless integrable hierarchies. After recalling the original
approach to the dKP hierarchy, we rederive dispersionless Hirota equations for the
dToda hierarchy and apply the method to the EdBKP hierarchy. The results enables us to
investigate the associativity equations in the EdBKP hierarchy.

Three remarks are in order.  Firstly, it would be interesting  to compare our results to
 a recent work by Takasaki\cite{T06} on two-component dBKP hierarchy where
 the corresponding dHirota equations were obtained by taking quasi-classical limit
 of the (differential) Fay identities. This will help us to understand
 how  to extend the dBKP hierarchy to  integrable hierarchies with two sets of
 time variables. Secondly, the associativity equations discussed in sec. 4 is
an infinite-dimensional version of WDVV equations in which the number of variables $t_k$
is infinite. However, in \cite{ChT}, some finite-dimensional reductions of the EdBKP hierarchy
have been obtained via Riemann-Hilbert construction.
Therefore, it is quite natural to ask whether these solutions satisfy
 the finite-dimensional version of WDVV equations.
Thirdly, in \cite{P04,P06a,P06b} Pavlov  discussed dHirota equations of dBKP
 system from the point of view of Egorov hydrodynamic chains and constructed associated
 solutions of WDVV equations. His approach in some sense gave a Hamiltonian approach
 to the dBKP system. Therefore it is also interesting to investigate Hamiltonian formulation
 to the EdBKP hierarchy which has the advantage of studying solutions of WDVV equations.
 Works in these directions are now in progress.

%%%%%%%%%%%%%%%%%%%%%%%%%%%%%%%%%%%%%%%%%%%%%%%%%%%%%%%%%%%%%%%%%
{\bf Acknowledgments\/}\\
We would like to thank Maxim Pavlov for his useful comments on the manuscript.
This work is supported by the National Science Council of Taiwan
under Grant  NSC94-2112-M-194-009(MHT).
%%%%%%%%%%%%%%%%%%%%%%%%%%%%%%%%%%%%%%%%%%%%%%%%%%%%%%%%%%%%%
%%%%%%%%%%%%%%%%%%%%%%%%%%%%%%%%%%%%%%%%%%%%%%%%%%%%%%%%%%%%%%%%%%%%%%%%%%%%%%%
\appendix
\section{dHirota for dToda}
%%%%%%%%%%%%%%%%%%%%%%%%%%%%%%%%%%%%%%%%%%%%%%%%%%%%%%%%%%%%

Equation (\ref{dTodaH-1}).
\bea
\mu^{-2}: &&\qquad
          \frac{1}{2}\mF_{02}-\frac{1}{2}\mF_{01}^2-\mF_{11}=0, \no\\
\mu^{-3}: &&\qquad
          \frac{2}{3}\mF_{03}-\frac{1}{6}\mF_{01}^3-\frac{1}{2}\mF_{01}\mF_{02}
          -\mF_{01}\mF_{11}-\mF_{12}=0,  \no\\
\mu^{-4}: &&\qquad
          \frac{3}{4}\mF_{04}-\frac{1}{24}\mF_{01}^4-\frac{1}{3}\mF_{01}\mF_{03}-\mF_{01}\mF_{12}
          -\frac{1}{8}\mF_{02}^2-\frac{1}{2}\mF_{02}\mF_{11}  \no\\
          &&\qquad -\frac{1}{2}\mF_{11}^2-\frac{2}{3}\mF_{13}-\frac{1}{4}\mF_{22}=0.
\label{dH-t1}
\eea
Equation (\ref{dTodaH-2}).
\bea
\mu^{-2}: &&\qquad
          \frac{1}{2}\mF_{0\bar{2}}+\frac{1}{2}\mF_{0\bar{1}}^2+\mF_{\bar{1}\bar{1}}=0, \no\\
\mu^{-3}: &&\qquad
          \frac{2}{3}\mF_{0\bar{3}}-\frac{1}{6}\mF_{0\bar{1}}^3+\frac{1}{2}\mF_{0\bar{1}}\mF_{0\bar{2}}
          -\mF_{0\bar{1}}\mF_{\bar{1}\bar{1}}+\mF_{\bar{1}\bar{2}}=0,  \no\\
\mu^{-4}: &&\qquad
          \frac{3}{4}\mF_{0\bar{4}}+\frac{1}{24}\mF_{0\bar{1}}^4+\frac{1}{3}\mF_{0\bar{1}}\mF_{0\bar{3}}
          -\mF_{0\bar{1}}\mF_{\bar{1}\bar{2}}+\frac{1}{8}\mF_{0\bar{2}}^2
          -\frac{1}{2}\mF_{0\bar{2}}\mF_{\bar{1}\bar{1}}  \no\\
          &&\qquad +\frac{1}{2}\mF_{\bar{1}\bar{1}}^2+\frac{2}{3}\mF_{\bar{1}\bar{3}}
           +\frac{1}{4}\mF_{\bar{2}\bar{2}}=0.
\label{dH-t2}
\eea
Equation (\ref{dTodaH-3}).
\bea
\mu^{-2}: &&\qquad
          e^{\mF_{00}}+\mF_{1\bar{1}}=0, \no\\
\mu^{-3}: &&\qquad
          e^{\mF_{00}}(\mF_{01}-\mF_{0\bar{1}})+\frac{1}{2}\mF_{1\bar{2}}
          +\frac{1}{2}\mF_{2\bar{1}}=0,  \no\\
\mu^{-4}: &&\qquad
          \frac{1}{2}e^{\mF_{00}}
          (\mF_{01}^2+\mF_{0\bar{1}}^2+\mF_{02}-\mF_{0\bar{2}}-2\mF_{01}\mF_{0\bar{1}}) \no\\
          &&\qquad
          +\frac{1}{2}\mF_{1\bar{1}}^2+\frac{1}{3}\mF_{1\bar{3}}\mF_{3\bar{1}}+\frac{1}{4}\mF_{2\bar{2}}=0.
\label{dH-t3}
\eea

%%%%%%%%%%%%%%%%%%%%%%%%%%%%%%%%%%%%%%%%%%%%%%%%%%%%%%%%%%%%%%%%%%%%%
\section{dHirota for EdBKP}
%%%%%%%%%%%%%%%%%%%%%%%%%%%%%%%%%%%%%%%%%%%%%%%%%%%%%%%%%%%%%%%%%%%%%%%%%%%%%%%%
Equation (\ref{Fmn-1}).
%%%%%%%%%%%%%%%%%%%%%%%%%%%%%%%%%%%%%%%%%%%%%%%%%%%%%%%%%%%%%%%%%%%%%%%%%%%%%%%
\bea
\mu^{-6}: &&\qquad
\frac{1}{5}\mF_{15}+\frac{1}{3}\mF_{11}^3-\frac{1}{3}\mF_{11}\mF_{13}-\frac{1}{9}\mF_{33}=0,\no\\
\mu^{-8}: &&\qquad
\frac{2}{7}\mF_{17}+\frac{1}{3}\mF_{11}^{4}+\frac{1}{3}\mF_{11}^2\mF_{13}
-\frac{1}{5}\mF_{11}\mF_{15}-\frac{1}{9}\mF_{11}\mF_{33}-\frac{2}{9}\mF_{13}^2-\frac{2}{15}\mF_{35}=0,\no\\
\mu^{-10}: &&\qquad
\frac{1}{3}\mF_{19}+\frac{1}{5}\mF_{11}^5+\frac{2}{3}\mF_{11}^3\mF_{13}+\frac{1}{5}\mF_{11}^2\mF_{15}
-\frac{1}{7}\mF_{11}\mF_{17}-\frac{2}{15}\mF_{11}\mF_{35} \no\\
&&\qquad\quad
-\frac{4}{15}\mF_{13}\mF_{15}-\frac{1}{9}\mF_{13}\mF_{33}-\frac{2}{21}\mF_{37}-\frac{1}{25}\mF_{55}=0.
\label{dHirota-eq1}
\eea
%%%%%%%%%%%%%%%%%%%%%%%%%%%%%%%%%%%%%%%%%%%%%%%%%%%%%%%%%%%%%%%%%%%%%%%%%%%%%%%%
Equation (\ref{Fmn-2}).
%%%%%%%%%%%%%%%%%%%%%%%%%%%%%%%%%%%%%%%%%%%%%%%%%%%%%%%%%%%%%%%%%%%%%%%%%%%%%%%
\bea
\bmu^{6}: &&\qquad
\frac{1}{5}\mF_{\bar{1}\bar{5}}+\frac{1}{3}\mF_{\bar{1}\bar{1}}^3
-\frac{1}{3}\mF_{\bar{1}\bar{1}}\mF_{\bar{1}\bar{3}}-\frac{1}{9}\mF_{\bar{3}\bar{3}}=0,\no\\
\bmu^{8}: &&\qquad
\frac{2}{7}\mF_{\bar{1}\bar{7}}+\frac{1}{3}\mF_{\bar{1}\bar{1}}^{4}
+\frac{1}{3}\mF_{\bar{1}\bar{1}}^2\mF_{\bar{1}\bar{3}}
-\frac{1}{5}\mF_{\bar{1}\bar{1}}\mF_{\bar{1}\bar{5}}-\frac{1}{9}\mF_{\bar{1}\bar{1}}\mF_{\bar{3}\bar{3}}
-\frac{2}{9}\mF_{\bar{1}\bar{3}}^2-\frac{2}{15}\mF_{\bar{3}\bar{5}}=0,\no\\
\bmu^{10}: &&\qquad
\frac{1}{3}\mF_{\bar{1}\bar{9}}+\frac{1}{5}\mF_{\bar{1}\bar{1}}^5
+\frac{2}{3}\mF_{\bar{1}\bar{1}}^3\mF_{\bar{1}\bar{3}}+\frac{1}{5}\mF_{\bar{1}\bar{1}}^2\mF_{\bar{1}\bar{5}}
-\frac{1}{7}\mF_{\bar{1}\bar{1}}\mF_{\bar{1}\bar{7}}-\frac{2}{15}\mF_{\bar{1}\bar{1}}\mF_{\bar{3}\bar{5}} \no\\
&&\qquad\quad
-\frac{4}{15}\mF_{\bar{1}\bar{3}}\mF_{\bar{1}\bar{5}}-\frac{1}{9}\mF_{\bar{1}\bar{3}}\mF_{\bar{3}\bar{3}}
-\frac{2}{21}\mF_{\bar{3}\bar{7}}-\frac{1}{25}\mF_{\bar{5}\bar{5}}=0.
\label{dHirota-eq2}
\eea
%%%%%%%%%%%%%%%%%%%%%%%%%%%%%%%%%%%%%%%%%%%%%%%%%%%%%%%%%%%%%%%%%%%%%%%%%%%%%%%%
Equation (\ref{Fmn-3}).
%%%%%%%%%%%%%%%%%%%%%%%%%%%%%%%%%%%%%%%%%%%%%%%%%%%%%%%%%%%%%%%%%%%%%%%%%%%%%%%
\bea
\mu^{-4}: &&\qquad
          \mF_{11}\mF_{1\bar{1}}-\frac{1}{3}\mF_{3\bar{1}}=0, \no\\
\mu^{-6}: &&\qquad
\mF_{11}\mF_{1\bar{1}}^2+\frac{1}{3}\mF_{11}\mF_{1\bar{3}}+\frac{1}{3}\mF_{11}\mF_{3\bar{1}}
+\frac{1}{3}\mF_{1\bar{1}}^3+\frac{1}{3}\mF_{1\bar{1}}\mF_{13}
-\frac{1}{3}\mF_{1\bar{1}}\mF_{3\bar{1}} \no\\
&&\qquad -\frac{1}{9}\mF_{3\bar{3}}-\frac{1}{5}\mF_{5\bar{1}}
=0,  \no\\
\mu^{-8}: &&\qquad
\frac{2}{3}\mF_{11}\mF_{1\bar{1}}\mF_{3\bar{1}}
+\frac{2}{3}\mF_{1 1}\mF_{1 \bar{1}}^3
+\frac{1}{5}\mF_{1 1}\mF_{1 \bar{5}}
+\frac{1}{9}\mF_{1 1}\mF_{3 \bar{3}}
+\frac{1}{5}\mF_{1 1}\mF_{5 \bar{1}} \no\\
&&\qquad
+\frac{2}{3}\mF_{1 1}\mF_{1 \bar{1}}\mF_{1 \bar{3}}
+\frac{1}{3}\mF_{1 3}\mF_{1 \bar{1}}^2
+\frac{1}{9}\mF_{1 3}\mF_{1 \bar{3}}
+\frac{1}{9}\mF_{1 3}\mF_{3 \bar{1}}
+\frac{1}{3}\mF_{1 \bar{1}}^4 \no\\
&&\qquad
+\frac{1}{3}\mF_{1 \bar{1}}^2\mF_{1 \bar{3}}
-\frac{1}{9}\mF_{1 \bar{1}}\mF_{3 \bar{3}}
-\frac{1}{5}\mF_{1 \bar{1}}\mF_{5 \bar{1}}
+\frac{1}{5}\mF_{1 \bar{1}}\mF_{1 5}
-\frac{1}{9}\mF_{1 \bar{3}}\mF_{3 \bar{1}} \no\\
&&\qquad
-\frac{1}{9}\mF_{3 \bar{1}}^2
-\frac{1}{15}\mF_{3 \bar{5}}
-\frac{1}{15}\mF_{5 \bar{3}}
-\frac{1}{7}\mF_{7 \bar{1}}
=0.
\label{dHirota-eq3}
\eea
%%%%%%%%%%%%%%%%%%%%%%%%%%%%%%%%%%%%%%%%%%%%%%%%%%%%%%%%%%%%%%%%%%%%%%%%%%%%%%%%
Equation (\ref{Fmn-4}).
%%%%%%%%%%%%%%%%%%%%%%%%%%%%%%%%%%%%%%%%%%%%%%%%%%%%%%%%%%%%%%%%%%%%%%%%%%%%%%%
\bea
\la^{-4}: &&\qquad
          \mF_{\bar{1}\bar{1}}\mF_{1\bar{1}}-\frac{1}{3}\mF_{1\bar{3}}=0, \no\\
\la^{-6}: &&\qquad
\mF_{\bar{1}\bar{1}}\mF_{\bar{1}1}^2+\frac{1}{3}\mF_{\bar{1}\bar{1}}\mF_{\bar{1}3}
+\frac{1}{3}\mF_{\bar{1}\bar{1}}\mF_{\bar{3}1}
+\frac{1}{3}\mF_{\bar{1}1}^3+\frac{1}{3}\mF_{\bar{1}1}\mF_{\bar{1}\bar{3}}
-\frac{1}{3}\mF_{\bar{1}1}\mF_{\bar{3}1} \no\\
&&\qquad -\frac{1}{9}\mF_{\bar{3}3}-\frac{1}{5}\mF_{\bar{5}1}
=0,  \no\\
\la^{-8}: &&\qquad
\frac{2}{3}\mF_{\bar{1}\bar{1}}\mF_{\bar{1}1}\mF_{\bar{3}1}
+\frac{2}{3}\mF_{\bar{1}\bar{1}}\mF_{\bar{1}1}^3
+\frac{1}{5}\mF_{\bar{1}\bar{1}}\mF_{\bar{1}5}
+\frac{1}{9}\mF_{\bar{1}\bar{1}}\mF_{\bar{3}3}
+\frac{1}{5}\mF_{\bar{1}\bar{1}}\mF_{\bar{5}1} \no\\
&&\qquad
+\frac{2}{3}\mF_{\bar{1}\bar{1}}\mF_{\bar{1} 1}\mF_{\bar{1}3}
+\frac{1}{3}\mF_{\bar{1}\bar{3}}\mF_{\bar{1}1}^2
+\frac{1}{9}\mF_{\bar{1}\bar{3}}\mF_{\bar{1}3}
+\frac{1}{9}\mF_{\bar{1}\bar{3}}\mF_{\bar{3}1}
+\frac{1}{3}\mF_{\bar{1}1}^4 \no\\
&&\qquad
+\frac{1}{3}\mF_{\bar{1}1}^2\mF_{\bar{1}3}
-\frac{1}{9}\mF_{\bar{1}1}\mF_{\bar{3}3}
-\frac{1}{5}\mF_{\bar{1}1}\mF_{\bar{5}1}
+\frac{1}{5}\mF_{\bar{1}1}\mF_{\bar{1}\bar{5}}
-\frac{1}{9}\mF_{\bar{1}3}\mF_{\bar{3}1} \no\\
&&\qquad
-\frac{1}{9}\mF_{\bar{3}1}^2
-\frac{1}{15}\mF_{\bar{3}5}
-\frac{1}{15}\mF_{\bar{5}3}
-\frac{1}{7}\mF_{\bar{7}1}
=0.
\label{dHirota-eq4}
\eea

%%%%%%%%%%%%%%%%%%%%%%%%%%%%%%%%%%%%%%%%%%%%%%%%%%%%%%%%%%%%%%

\end{document}